\documentclass[aps,prc,twocolumn,superscriptaddress,nofootinbib]{revtex4-1}
\usepackage{latexsym} 
\usepackage{amsmath}
\usepackage{amssymb}
\usepackage{amsfonts}
\usepackage{bm}
\usepackage{physics}
\usepackage{booktabs}

\usepackage{color}
\definecolor{purple}{rgb}{0.5,0,0.5}
\definecolor{blue}{rgb}{0.0,0,0.9}
\definecolor{prdblue}{rgb}{0.133,0.118,0.498}
\usepackage[colorlinks=true, pdfstartview=FitV, linkcolor=prdblue, citecolor= prdblue, urlcolor=prdblue]{hyperref}

\usepackage{supertabular}
\usepackage{placeins}
\usepackage{epsfig}
\usepackage{graphicx}

\usepackage{CJKutf8}

\usepackage{soul}

\begin{document}
\begin{CJK*}{UTF8}{gbsn}

\title{A semi-classical study of muon-enhanced proton-boron-11 fusion}

\author{Hong-Yi Wang (王弘毅)}
\affiliation{School of Physics, Nanjing University, Nanjing, Jiangsu 210093, China}

\author{Ming-Yu Chen (陈铭宇)}
\affiliation{School of Physics, Nanjing University, Nanjing, Jiangsu 210093, China}

\author{Hao-Le Ma (马好乐)}
\affiliation{School of Physics, Nanjing University, Nanjing, Jiangsu 210093, China}

\author{Zhu-Fang Cui (崔著钫)}
\email[Contact author: ]{phycui@nju.edu.cn}
\affiliation{School of Physics, Nanjing University, Nanjing, Jiangsu 210093, China}

\date{\today}

\begin{abstract}
A recent theoretical study has suggested that muons can enhance proton-boron-11 (p-$^{11}$B) reaction cross-section by several orders of magnitude in the low-energy regime. In this work, we investigate this reaction process using a semi-classical treatment, that is, a muon and a proton first form a muonic hydrogen atom p$\mu$, which subsequently collides with a $^{11}$B nucleus. During the collision, the p$\mu$ atom approaches the $^{11}$B nucleus and is then repelled by the Coulomb repulsive potential. At the distance of closest approach between the proton and the $^{11}$B nuclei in this classical scattering process---namely, the classical turning point---quantum tunneling through the Coulomb barrier can occur, allowing the proton to penetrate into the range of the nuclear force of the $^{11}$B and trigger the fusion reaction. We determine the turning point statistically by using the classical trajectory Monte Carlo method, where the initial phase-space distributions of the proton and muon are sampled from the ground-state microcanonical distribution. Our results show that, compared with the bare-nucleus case, the reaction cross-section is enhanced by several orders of magnitude in the low-energy region. A comparison with the static charge-shielding treatment reveals certain differences; however, both approaches demonstrate that the catalytic effect of the muon can significantly enhance the low-energy p-$^{11}$B reaction cross-section.
\end{abstract}

\maketitle
\end{CJK*}


\section{Introduction}
Proton--boron (p-$^{11}$B) fusion is an attractive aneutronic fusion reaction pathway that has received renewed interest in recent years~\cite{Last2011,Liu2024,Sciscio2025}. The reaction,
\[
\mathrm{p} + {}^{11}\mathrm{B} \rightarrow 3\alpha + 8.7\ \mathrm{MeV},
\]
converts the reactants entirely into three charged $\alpha$-particles without producing high-energy neutrons~\cite{Beckman1953,Moreau1977}. This reaction offers three principal advantages: first, both reactants are abundant and stable in nature, avoiding the formidable radiological handling issues~\cite{Liu2024,Mazzucconi2025}; second, its aneutronic nature eliminates the material activation and nuclear-waste management challenges inherent in conventional deuterium-tritium (D--T) fusion~\cite{Dawson1981,Ogawa_2024}; and third, the resulting charged $\alpha$-particles permit direct energy conversion, offering a significant efficiency gain over neutron-dominated approaches~\cite{Moreau1977,Liu2024}. Despite these appealing features, the practical feasibility of p-$^{11}$B fusion has long been questioned. Owing to its relatively low reactivity, ignition requires ion temperatures on the order of several hundred keV, significantly higher than those for D--T fusion~\cite{Putvinski2019,Hartouni_2022,Tentori2023}. Therefore, developing methods to enhance the p-$^{11}$B reactivity has become a crucial objective in p-$^{11}$B fusion research~\cite{Wang:2026zuj}.

Muon-catalyzed fusion (MCF or $\mu$CF) represents one promising strategy for enhancing reactivity, and the application of MCF to D--T fusion has already been extensively investigated both theoretically and experimentally. In the D--T case, the presence of a muon enables the formation of a $\mathrm{DT}\mu$ molecule, thereby realizing so-called cold fusion, or promotes fusion through the in-flight fusion mechanism~\cite{Rafelski1987,Froelich1992,Wu2024,Wu:2025bnm}. However, the application of MCF to p-$^{11}$B fusion has been scarcely explored. This is not only because of the relatively high atomic number of the $^{11}$B ($\mathbb{Z}=5$) prevents the formation of a quasi-bound $\mathrm{B}$p$\mu$ state, but also because the $^{11}$B nucleus itself acts as an efficient muon trap: it can capture a muon, bind it tightly, and potentially induce the decay of the $^{11}$B to $^{11}$Be~\cite{Koshigiri1984,Schaller1993,PhysRevC.65.025503}. Only very recently has a feasible scheme been proposed, namely, a muon first binds with a proton to form a relatively stable muonic hydrogen atom p$\mu$, which is then bombarded by a $^{11}$B nucleus~\cite{Wang:2026zuj}. Owing to the unit negative charge of the muon, one unit of positive charge of the proton is effectively screened; moreover, since the muon mass is approximately 207 times of the electron, the Bohr radius of p$\mu$ is about 207 times smaller than that of ordinary hydrogen. This results in a much more tightly bound muon cloud and a significantly increased charge-screening range. Although this proposal offers an important catalytic pathway, the cross-section calculation in that work assumes that the muon remains strictly in the ground state throughout the entire reaction process, unaffected by the $^{11}$B nucleus, and reduces the effect of the muon to a distance-dependent modification of the proton's effective charge, thereby treating the problem as a two-body p--$^{11}$B system. While such a treatment is intuitively appealing and computationally convenient, its limitations are evident: first, given the strong attractive force exerted by the $^{11}$B on the muon, the assumption that the muon remains in the static ground state is inaccurate, as the muon trajectory will inevitably be distorted by the $^{11}$B nucleus; second, simplifying the p$\mu$--$^{11}$B system to a two-body problem with a dynamic charge distribution neglects the genuine three-body interactions, thus compromising its ability to describe the real physical process.

To simulate the reaction more realistically, we propose a semi-classical (SC) treatment in which a muon and a proton first form a muonic hydrogen atom p$\mu$, which subsequently collides with a $^{11}$B nucleus. During the collision, the p$\mu$ atom and the $^{11}$B nucleus approach each other and then separate again under the Coulomb repulsive potential. At the distance of closest approach between the p and the $^{11}$B nuclei in this classical scattering process---namely, the classical turning point---quantum tunneling through the Coulomb barrier can occur, allowing the nuclei to penetrate into the range of the nuclear force and trigger the fusion reaction. Because the timescale of the nuclear interaction is much shorter than that of the electromagnetic interaction, we assume that once tunneling occurs and the nuclei enter the nuclear force range, the fusion reaction takes place instantaneously. The turning-point distance is determined statistically by using the classical trajectory Monte Carlo (CTMC) method~\cite{percival1975theory,liu2013classical}, where the initial phase-space distributions of the proton and muon are sampled from the microcanonical ensemble at the ground-state energy. We employ the Kepler equation of planetary motion to randomly sample the initial conditions of the p$\mu$ atom, generating a large ensemble of initial configurations~\cite{abrines1966classical}. We verify that the momentum distribution obtained from this Kepler sampling agrees with that derived from the quantum-mechanical ground-state wave function, confirming that our random initial states satisfy the intended microcanonical distribution. We write down the full three-body Hamiltonian of the p$\mu$--$^{11}$B system and use Hamilton's canonical equations to compute the trajectory for each sampled initial state, from which we extract the minimum p--$^{11}$B distance along each trajectory. The equations of motion are integrated using the Velocity-Verlet algorithm, which provides high-accuracy trajectory tracing over long simulation times while preserving excellent energy conservation~\cite{swope1982computer,verlet1967computer}. By averaging over a large number of trajectories for a given incident energy, we obtain the mean turning-point distance as a function of energy---that is, the position at which tunneling is initiated for different incident energies. The tunneling probability at each turning point is evaluated using the WKB approximation when the action $S \geq 10$, and is connected smoothly by the Airy approximation when $S < 10$. Finally, the reaction cross-section is obtained from the Gamow form.

This paper is organized as follows. Sec.~\ref{SCM} presents our semi-classical treatment including the three-body Hamiltonian, the CTMC method and the penetrability. Sec.~\ref{results} gives the results and discussion. Conclusions are given in Sec.~\ref{conclusion}.


\section{Framework}
\label{SCM}
\subsection{Hamiltonian}
\begin{figure}
    \centering
    \includegraphics[width=\linewidth]{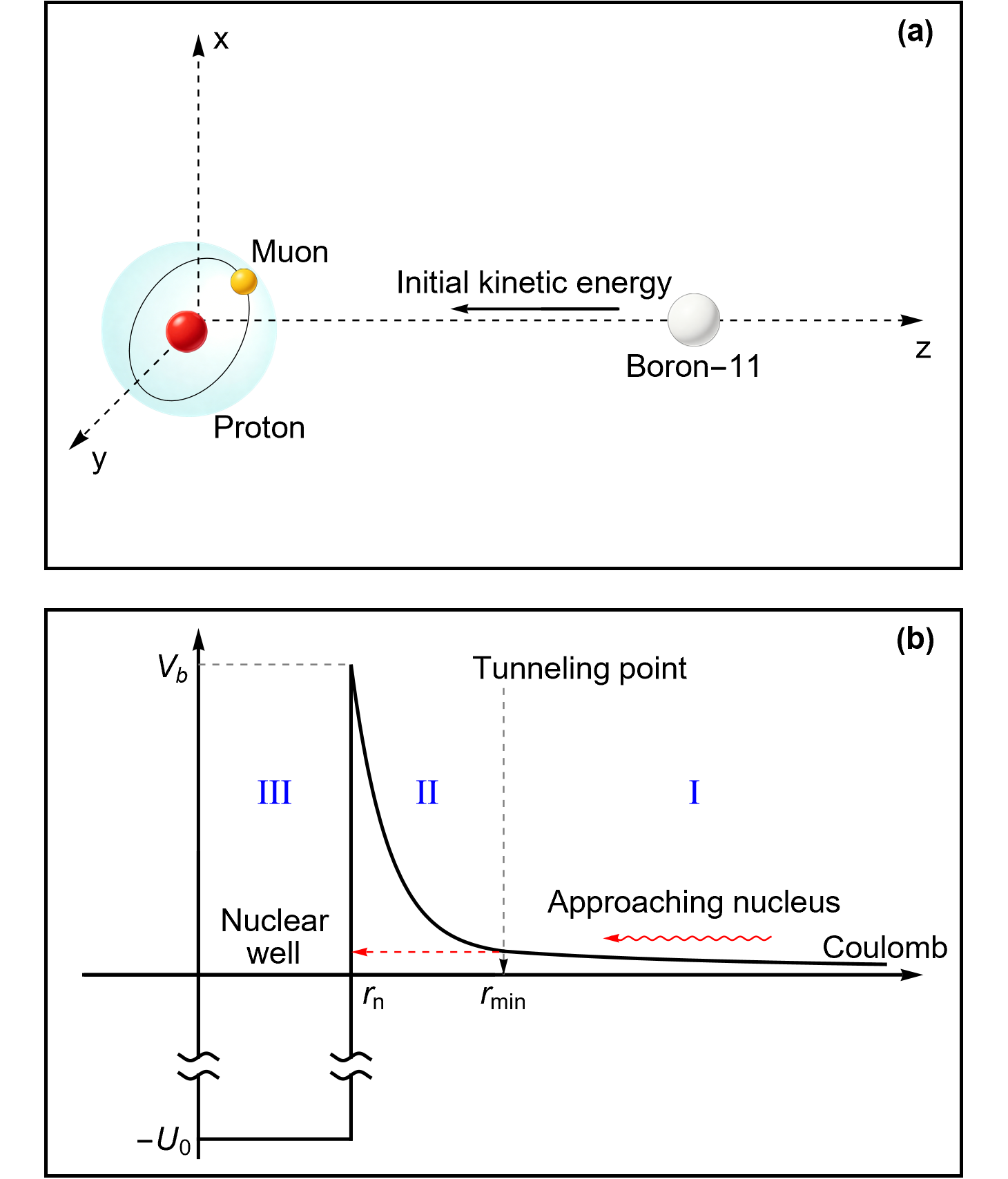}
    \caption{(Color online) (a) Schematic description of a muonic atom $p\mu$ impacted by a $^{11}$B. (b) The process of p$\mu$--$^{11}$B fusion is divided into three regions. Region I: classical motion; Region II: quantum tunneling; Region III: nuclear fusion.}
    \label{fig:SC_process}
\end{figure}

We consider a three-body p$\mu$--$^{11}$B system, as sketched in Fig.~\ref{fig:SC_process}(a). For computational convenience, the initial position of the p$\mu$ atom is placed at the origin of the coordinate system; i.e., the center of mass of the p$\mu$ atom is set as the coordinate origin. The $^{11}$B nucleus is initially placed on the $z$-axis far from the origin and is given an initial kinetic energy directed toward the p$\mu$ atom along the $z$-axis. This coordinate choice facilitates both the initial-state sampling and the dynamical integration. However, all initial kinetic energies and calculated results mentioned in this paper are expressed in the center-of-mass frame for consistency and brevity. The muon, proton, and $^{11}$B nucleus are all treated as point particles. This approximation is justified by the fact that the de Broglie wavelength of each particle, $\lambda = 2\pi\hbar/\sqrt{2m\varepsilon}$, is much smaller than the characteristic interaction length for incident kinetic energies below $20$~MeV---where the interaction length can be estimated by the distance over which the Coulomb potential changes by $1\%$. The energy range considered in this work fully satisfies this condition. A schematic illustration of the potential experienced by the $^{11}$B nucleus as it approaches the origin is shown in Fig.~\ref{fig:SC_process}(b). The entire process is divided into three regions: in Region~I, the three bodies undergo classical motion, and the p--$^{11}$B nuclei reach the distance of closest approach $r_{\min}$ under the influence of the Coulomb repulsive potential; in Region~II, quantum tunneling occurs from $r_{\min}$ to $r_{\mathrm{n}}$ ($r_{\mathrm{n}} = 3.284$~fm is the contact radius of p--$^{11}$B nuclei~\cite{IAEANuclearData}); the nuclei then enter the range of the nuclear force (Region~III) and the fusion reaction takes place.

The Hamiltonian of this three-body system can be written as
\begin{equation}
    \begin{aligned}
        H = & \frac{\bm{p}_{\mathrm{p}}^2}{2m_{\mathrm{p}}} + \frac{\bm{p}_{\mathrm{B}}^2}{2m_{\mathrm{B}}} + \frac{q_{\mathrm{p}}q_{\mathrm{B}}e^2}{4\pi \varepsilon_0 |\bm{r}_{\mathrm{p}} - \bm{r}_{\mathrm{B}}|} \\
        & + \frac{\bm{p}_{\mu}^2}{2m_{\mu}} - \frac{q_{\mu}q_{\mathrm{p}}e^2}{4\pi \varepsilon_0 |\bm{r}_{\mu} - \bm{r}_{\mathrm{p}}|} - \frac{q_{\mu}q_{\mathrm{B}}e^2}{4\pi \varepsilon_0 |\bm{r}_{\mu} - \bm{r}_{\mathrm{B}}|},
    \end{aligned}
\end{equation}
where $m_i$ and $q_i$ ($i = \mathrm{p}$, $\mathrm{B}$, $\mu$ denoting the proton, $^{11}$B nucleus, and muon, respectively) are the masses and charges of the particles, specified by the position vectors $\bm{r}_i$ and momentum vectors $\bm{p}_i$, and $\varepsilon_0$ is the vacuum permittivity.

The first Bohr radius and the ground-state energy of the p$\mu$ atom can be estimated as $a_{\mu} = 4\pi\varepsilon_0\hbar^{2}/(m_{\mu}e^{2}) \approx 284.6$~fm and $E_{\mathrm{g.s.}} = -m_{\mu} e^{4} /[(4\pi\varepsilon_0)^{2} 2\hbar^{2}] \approx -2.815$~keV, respectively~\cite{Wang:2026zuj}. In Region~I, the classical motions are governed by the following canonical equations:
\begin{equation}\label{cequation}
    \frac{d\bm{r}_i}{dt} = \frac{\partial H}{\partial \bm{p}_i}, \quad \frac{d\bm{p}_i}{dt} = -\frac{\partial H}{\partial \bm{r}_i}.
\end{equation}

\subsection{Classical trajectory Monte Carlo}
\label{ctmc}
\begin{figure}
    \centering
    \includegraphics[width=\linewidth]{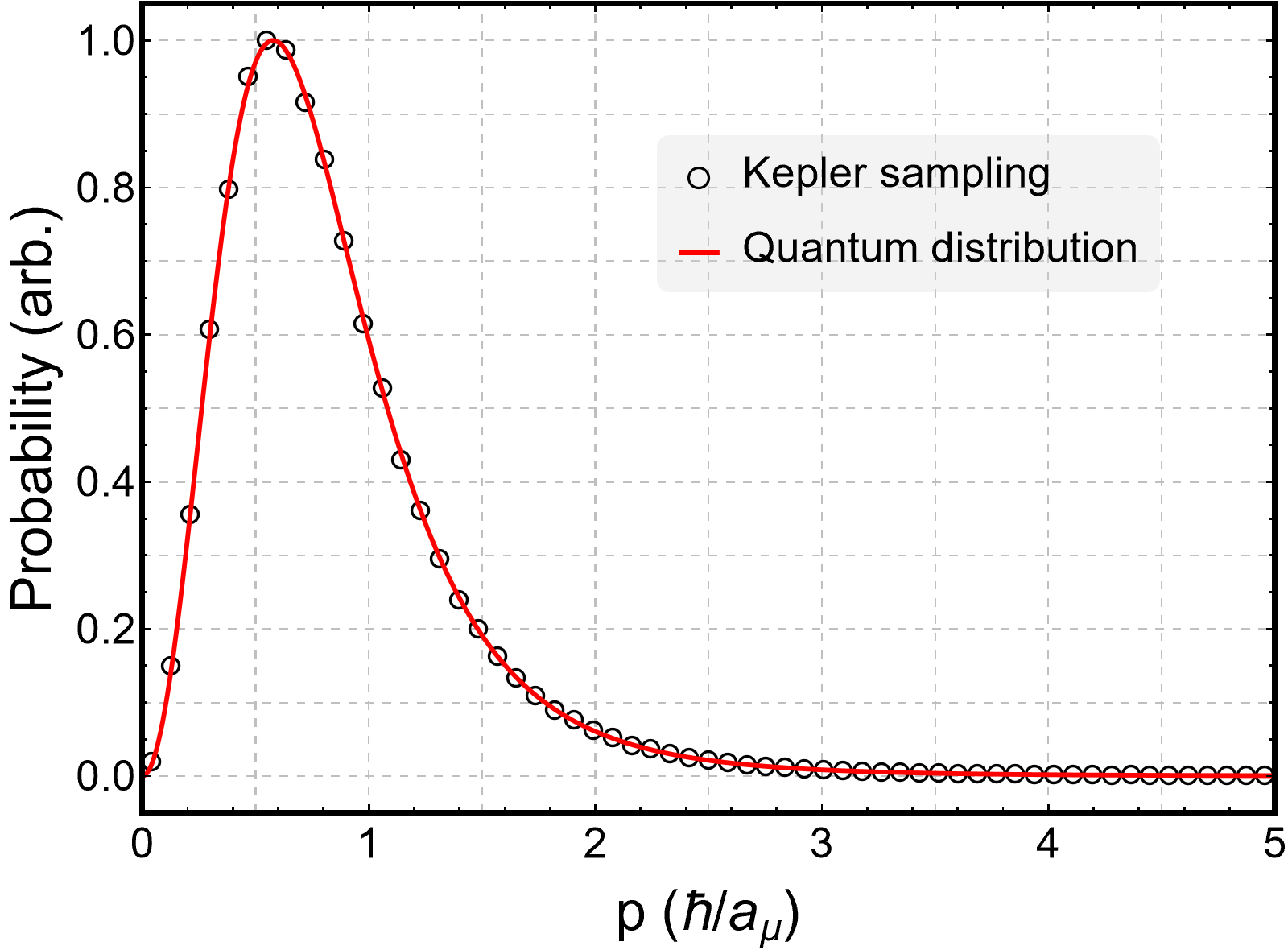}
    \caption{(Color online) Comparison of the momentum distribution obtained from Kepler sampling (black circle) with the exact quantum-mechanical ground-state distribution (red line).}
    \label{fig:Kepler_distribution}
\end{figure}

To determine the classical turning-point distance from the Hamiltonian for a given initial kinetic energy, we employ the classical trajectory Monte Carlo (CTMC) method~\cite{abrines1966classical,percival1975theory,liu2013classical}. This requires specifying physically realistic initial conditions --- the initial positions and momenta of the proton, $^{11}$B, and muon. In the coordinate system defined in Fig.~\ref{fig:SC_process}, the initial state of the $^{11}$B nucleus is straightforwardly given by
\begin{equation}
    \begin{aligned}
        \bm{r}_{\mathrm{B}}^{\mathrm{ini.}} &= [0,\, 0,\, z_{\mathrm{B}}^{\mathrm{ini.}}], \\[2pt]
        \bm{p}_{\mathrm{B}}^{\mathrm{ini.}} &= [0,\, 0,\, -m_{\mathrm{B}} \cdot v_{\mathrm{B}}^{\mathrm{ini.}}],
    \end{aligned}
\end{equation}
where $v_{\mathrm{B}}^{\mathrm{ini.}} = \sqrt{2\,\frac{m_{\mu}+m_{\mathrm{p}}+m_{\mathrm{B}}}{(m_{\mu}+m_{\mathrm{p}})\,m_{\mathrm{B}}}\,E}$. Since the $^{11}$B nucleus is incident from a macroscopic distance, we set $z_{\mathrm{B}}^{\mathrm{ini.}}$ far beyond the range where the Coulomb potential is appreciable; specifically, we use $z_{\mathrm{B}}^{\mathrm{ini.}} = 10^{6}$~fm throughout this work.

For the p$\mu$ atom, the initial state is sampled from the microcanonical distribution at the ground-state energy:
\begin{equation}\label{eq:microcanonical}
    \rho(\bm{r}, \bm{p}) \propto \delta\bigl[E_{\mathrm{g.s.}} - H_{\mathrm{p}\mu}(\bm{r}, \bm{p})\bigr],
\end{equation}
where $H_{\mathrm{p}\mu} = \bm{p}^{2}/2m_{\rho} - q_{\mu}q_{\mathrm{p}}e^{2}/(4\pi\varepsilon_{0}|\bm{r}|)$ is the relative-motion Hamiltonian of the p$\mu$ atom, $m_{\rho} = m_{\mu}m_{\mathrm{p}}/(m_{\mu}+m_{\mathrm{p}})$ is the reduced mass, and $\bm{r} = \bm{r}_{\mu} - \bm{r}_{\mathrm{p}}$, $\bm{p} = |m_{\mathrm{p}}\,\bm{p}_{\mu} - m_{\mu}\,\bm{p}_{\mathrm{p}}|/(m_{\mathrm{p}}+m_{\mu})$ are the relative position and momentum, respectively. Directly constructing the distribution from Eq.~\eqref{eq:microcanonical} is notoriously difficult. We therefore adopt a practical alternative: parameterize the p$\mu$ initial state using the Kepler equation of planetary motion and sample in the space of Kepler orbital parameters, which is equivalent to uniform sampling of the phase-space representative points~\cite{abrines1966classical,abrines1966classical2}.

Concretely, we first assume that the p$\mu$ atom moves in the $x$--$y$ plane on a Kepler orbit and express the relative coordinates as
\[
    \bm{r}_{x\text{-}y}^{\mathrm{ini.}} = \bigl[a\sqrt{1-\chi^{2}}\,\sin u,\; a(\cos u - \chi),\; 0\bigr],
\]
and the relative momenta as
\[
    \bm{p}_{x\text{-}y}^{\mathrm{ini.}} = \bigl[b\sqrt{1-\chi^{2}}\,\cos u/(1-\chi\cos u),\; -b\sin u/(1-\chi\cos u),\; 0\bigr],
\]
where $a = 1/(2E_{\mathrm{g.s.}})$, $b = \sqrt{2E_{\mathrm{g.s.}}}$, and $\chi$ and $u$ are the orbital eccentricity and the eccentric angle, respectively~\cite{cohen1982comment,leopold1979ionisation}. By drawing $\chi^{2}$ uniformly from $[0,1]$ and $u$ uniformly from $[0,2\pi]$, we obtain a two-dimensional random ensemble of initial states in the $x$--$y$ plane. To extend this to the three-dimensional geometry shown in Fig.~\ref{fig:SC_process}(a), we apply a random three-dimensional rotation to the two-dimensional initial states. The resulting initial conditions for the muon and the proton are
\begin{equation}\label{eq:Kepler_sampling}
    \begin{aligned}
        \bm{r}_{\mu}^{\mathrm{ini.}} &= \frac{m_{\mathrm{p}}}{m_{\mathrm{p}}+m_{\mu}}\,\mathbf{A}\,\bm{r}_{x\text{-}y}^{\mathrm{ini.}}, \\
        \bm{r}_{\mathrm{p}}^{\mathrm{ini.}} &= -\frac{m_{\mu}}{m_{\mathrm{p}}+m_{\mu}}\,\mathbf{A}\,\bm{r}_{x\text{-}y}^{\mathrm{ini.}}, \\
        \bm{p}_{\mu}^{\mathrm{ini.}} &= -\bm{p}_{\mathrm{p}}^{\mathrm{ini.}} = \mathbf{A}\,\bm{p}_{x\text{-}y}^{\mathrm{ini.}},
    \end{aligned}
\end{equation}
where $\mathbf{A}$ is a random three-dimensional Euler rotation matrix.

To verify that the Kepler sampling faithfully reproduces the required microcanonical distribution, we compare the momentum distribution obtained from Eq.~\eqref{eq:Kepler_sampling} with the exact quantum-mechanical momentum distribution of the muon in the ground state~\cite{LandauLifshitz1977}, which reads
\begin{equation}
    \rho_{\mathrm{Q.M.}}(\bm{p}) = \frac{8\,p_{c}^{2}}{\pi^{2}\,(p^{2} + p_{c}^{2})^{4}}, \qquad p_{c}^{2} = 2\,m_{\rho}\,|E_{\mathrm{g.s.}}|.
\end{equation}
As shown in Fig.~\ref{fig:Kepler_distribution}, the Kepler-sampled momentum distribution agrees perfectly with the quantum-mechanical result, confirming that our sampling scheme indeed generates the microcanonical ensemble described by Eq.~\eqref{eq:microcanonical}.

\begin{figure}
    \centering
    \includegraphics[width=\linewidth]{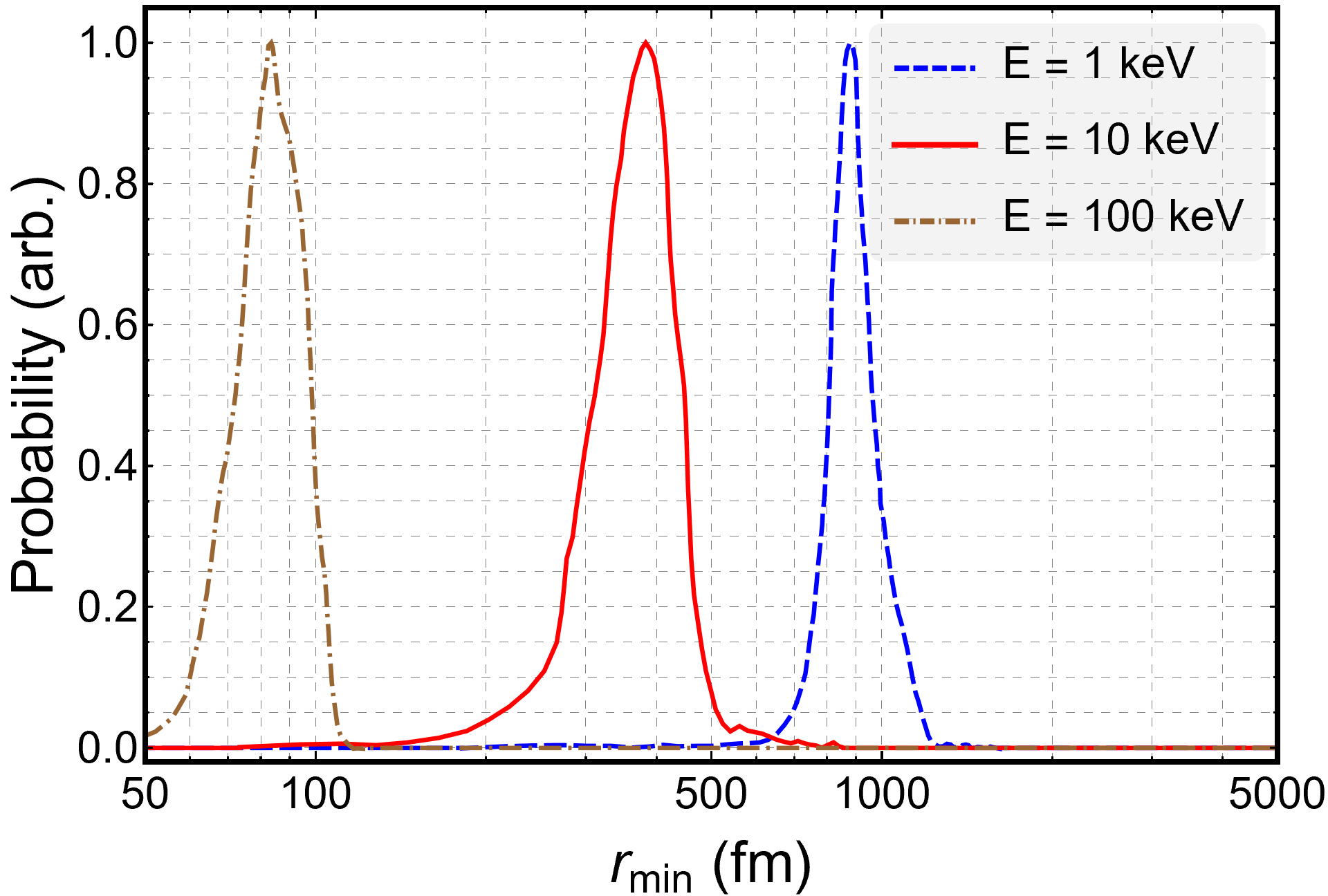}
    \caption{(Color online) Distribution of the classical turning-point distance $r_{\min}$ for incident energies $E = 1$, $10$, and $100$~keV. $10^{5}$ trajectories are taken for each energy.}
    \label{fig:CTMC_distribution}
\end{figure}

Once the initial conditions for all three particles are determined, we can integrate Eq.~(\ref{cequation}) numerically using the Velocity-Verlet algorithm and record the minimum p--$^{11}$B separation along each trajectory~\cite{swope1982computer,verlet1967computer}. For a fixed incident energy, this procedure is repeated for a large number of random initial states, yielding the statistical distribution of the turning-point distance. The average value is taken as the representative turning-point distance at that energy. Figure~\ref{fig:CTMC_distribution} shows typical turning-point distributions for $E = 1$, $10$, and $100$~keV. In all cases, the distribution forms a sharp peak around the mean value, indicating that the turning points obtained from different initial configurations are tightly clustered and that the ensemble average reliably characterizes the overall behavior.

\subsection{Penetrability}
\label{pene}
To calculate the tunneling probability, we use the WKB approximation supplemented by the Airy approximation in the regime where the WKB result becomes unreliable. We start from the radial Schr\"{o}dinger equation for the relative motion of the p--$^{11}$B system~\cite{LandauLifshitz1977},
\begin{equation}
\frac{d^{2} u}{d r^{2}}+k^{2}(r)\,u(r)=0, \qquad k^{2}(r) \equiv \frac{2\mu_{pB}}{\hbar^{2}}\bigl[E-V(r)\bigr],
\end{equation}
where $\mu_{pB}$ is the reduced mass of the proton and the $^{11}$B, and $V(r)$ is the Coulomb potential between them. In the classically forbidden region, the WKB wavefunction takes the form
\begin{equation}
u_{\mathrm{WKB}}(r) = \frac{C}{\sqrt{|k(r)|}}\,
\exp\!\left(-\int_{r_{\mathrm{n}}}^{r}\!\bigl|k(r^{\prime})\bigr|\,dr^{\prime}\right),
\end{equation}
with $C$ a normalization constant. From this, the WKB tunneling probability follows as
\begin{equation}\label{eq:PWKB}
P_{\mathrm{WKB}}=
\exp\!\left\{-\frac{2}{\hbar}\int_{r_{\mathrm{n}}}^{r_{\min}}\!
\sqrt{2m\bigl[V(r)-V(r_{\min})\bigr]}\;dr\right\}.
\end{equation}

The WKB formula~\eqref{eq:PWKB} is accurate only when the corresponding action $S \equiv -\frac{1}{2}\,\ln P_{\mathrm{WKB}}$ satisfies $S \gg 1$. Following common practice, we regard the WKB approximation as reliable for $S \geq 10$; for $S < 10$ we switch to the Airy-function connection technique~\cite{LandauLifshitz1977,Olver1974}. At the turning point $r_{t}$ where $S = 10$, the Coulomb potential is linearized,
\begin{equation}
V(r) \simeq E + V^{\prime}(r_{t})\,(r - r_{t}),
\end{equation}
and the Airy variable
\begin{equation}
\xi = \kappa\,(r - r_{t}),\qquad 
\kappa = \left(\frac{2\mu\,V^{\prime}(r_{t})}{\hbar^{2}}\right)^{\!1/3}
\end{equation}
is introduced. The radial equation then reduces to the standard Airy equation, whose general solution is $u_{\mathrm{Airy}}(\xi)=A\,\mathrm{Ai}(\xi)+B\,\mathrm{Bi}(\xi)$, with $\mathrm{Ai}$ and $\mathrm{Bi}$ the Airy functions~\cite{AbramowitzStegun1964}. The coefficients $A$ and $B$ are fixed by matching the Airy solution and its derivative to the WKB wavefunction and its derivative at a chosen matching point $r_{m}$:
\begin{equation}
\begin{bmatrix}
\mathrm{Ai}(\xi_{m}) & \mathrm{Bi}(\xi_{m}) \\[2pt]
\kappa\,\mathrm{Ai}^{\prime}(\xi_{m}) & \kappa\,\mathrm{Bi}^{\prime}(\xi_{m})
\end{bmatrix}
\begin{bmatrix}
A \\ B
\end{bmatrix}
=
\begin{bmatrix}
u_{\mathrm{WKB}}(r_{m}) \\[2pt]
u_{\mathrm{WKB}}^{\prime}(r_{m})
\end{bmatrix}.
\end{equation}
We take $r_{m}$ such that $|\xi_{m}| = 1$; at this point both the WKB and the Airy descriptions are valid, guaranteeing a smooth and physically consistent matching between the two approximations.

\section{Results}
\label{results}
\subsection{Minimum Distance and Penetrability}

\begin{figure*}
    \centering
    \includegraphics[width=\linewidth]{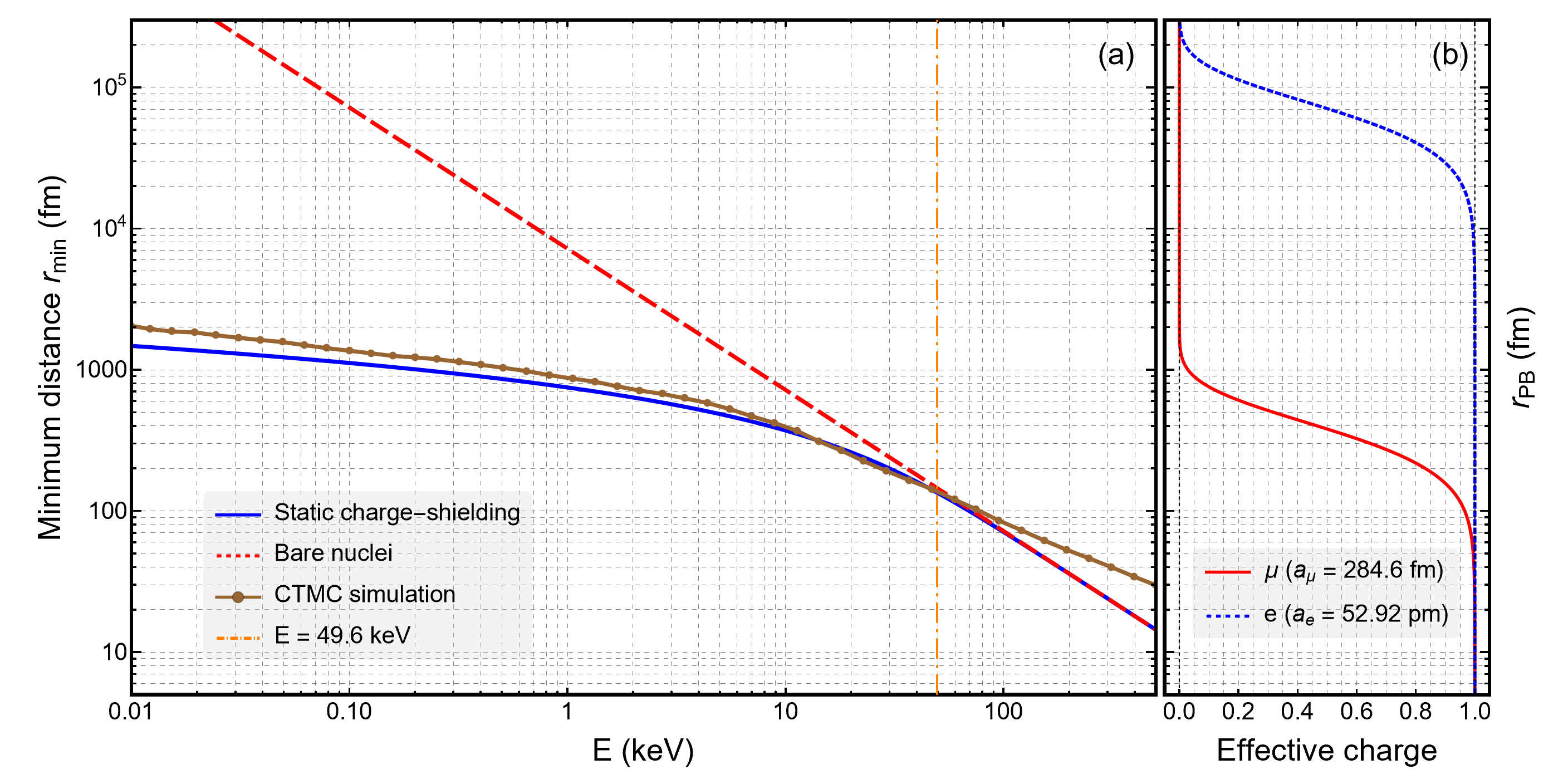}
    \caption{(Color online) (a) Representative turning-point distance $r_{\min}$ between the proton and the $^{11}$B as a function of the incident kinetic energy $E$. The brown curve shows the full CTMC result, the blue solid line corresponds to the static charge-shielding treatment, and the red dotted line represents the bare p--$^{11}$B case. (b) The total effective charge experienced by the $^{11}$B at a separation $r_{\rm pB}$ from the proton corresponding to the static charge-shielding treatment.}
    \label{fig:rvsE}
\end{figure*}

Using the numerical procedure described in Sec.~\ref{ctmc}, we have determined the turning-point distance between the proton and the $^{11}$B for incident kinetic energies ranging from $0.01$ to $500$~keV. For each energy, $10^{5}$ trajectories with randomly sampled initial conditions were propagated, and the turning-point distances were averaged to obtain a statistically robust result. The CTMC results are shown as the brown curve in Fig.~\ref{fig:rvsE}; for comparison, we also plot the turning-point distance for bare p--$^{11}$B nuclei and the result obtained from the static charge-shielding treatment as suggested in Ref.~\cite{Wang:2026zuj}.

In the energy range $E \approx 1$--$50$~keV, the CTMC results agree well with the static charge-shielding predictions. For $E \gtrsim 50$~keV, however, the two curves progressively diverge: the CTMC turning-point distance becomes systematically larger, and at $500$~keV it is approximately a factor of two larger than the static-treatment value. In the low-energy region $E \lesssim 1$~keV, the CTMC results also lie systematically above the static charge-shielding curve; the deviation grows gradually with decreasing energy, although it remains within $50\%$ over the entire sub-keV range.

These systematic discrepancies indicate that the static charge-shielding treatment --- which reduces the three-body p$\mu$--$^{11}$B system to a head-on collision between two point particles with a muon-modified effective charge on the proton --- provides accurate turning-point distances only over a limited energy window. At both lower and higher energies it systematically underestimates $r_{\min}$, and would therefore overestimate the reaction cross-section. The origin of this error can be understood as follows. The static screening picture implicitly assumes that the characteristic velocity of the muon is much larger than that of the incident $^{11}$B nucleus, so that the muon appears as a stationary charge cloud to the $^{11}$B. This condition breaks down when the $^{11}$B velocity becomes comparable to the muon's orbital velocity. To estimate the relevant energy scale, we note that the first Bohr radius of the p$\mu$ atom is $a_{\mu} \approx 284.6$~fm and the corresponding Bohr velocity is $\alpha c$; the incident kinetic energy at which the $^{11}$B nucleus reaches this velocity at $r = a_{\mu}$ is approximately $49.6$~keV. Above this energy, the static treatment predicts a turning-point distance nearly identical to the bare-nucleus result, simply because the effective charge of the proton is already close to unity and the muon screening becomes negligible. The CTMC simulation, in contrast, yields a turning-point distance even larger than the bare-nucleus value: once the $^{11}$B moves sufficiently fast, it not only fails to sense the muon as a screening cloud, but also experiences an additional attraction from the muon, which pulls the $^{11}$B slightly outward and increases $r_{\min}$.

The deviation in the low-energy regime ($E \lesssim 1$~keV) has a different origin. Although the incident $^{11}$B velocity is well below the muon's Bohr velocity, and the p--$^{11}$B separation remains in the region where charge screening is effective, the static treatment describes the collision as a pure head-on encounter between two point particles, thus ignoring the internal motion of the p$\mu$ atom. In our CTMC calculation, the center of mass of the p$\mu$ atom --- rather than the proton itself --- is placed at the origin (see Fig.~\ref{fig:SC_process}(a)). This naturally introduces a small but non-negligible impact parameter between the proton and the $^{11}$B, which systematically increases the turning-point distance compared with the head-on limit. This effect is more pronounced at low energies where it is more sensitive to any non-zero impact parameter.

\begin{figure}
	\centering
	\includegraphics[width=\linewidth]{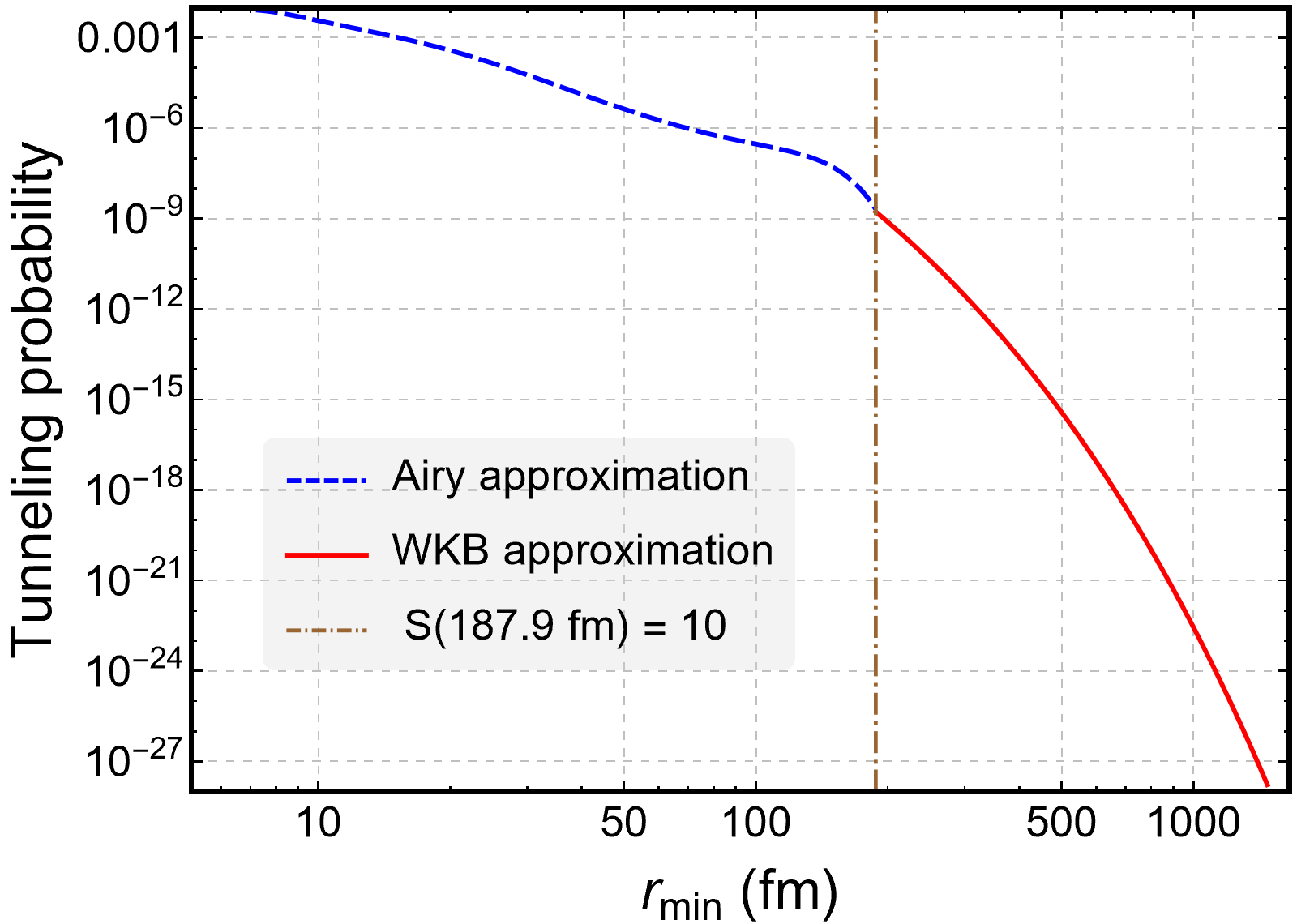}
    \caption{(Color online) Tunneling probability as a function of turning-point distance $r_{\min}$.} 
    \label{penetrability}
\end{figure}

The tunneling probability is computed following the procedure described in Sec.~\ref{pene}, and the results are presented in Fig.~\ref{penetrability}.  For turning-point distances $r_{\min} \ge 187.9$~fm, the corresponding action satisfies $S \ge 10$, and we employ the WKB approximation.  At $r_{\min} = 187.9$~fm, where $S = 10$, the Airy-function connection technique is applied to ensure a smooth transition; for smaller $r_{\min}$ (i.e., higher incident energies), the penetrability is obtained from the Airy approximation.

\subsection{Cross-Section and Reactivity}

The reaction cross-section serves as a crucial link between theory and experiment: experiments cannot directly measure the tunneling probability, but they can measure the cross-section.  The cross-section is customarily expressed in the Gamow form
\begin{equation}
\sigma(E)=\frac{S(E)}{E}\,P(E),
\end{equation}
where the factor $1/E$ reflects the geometrical effect of the incident wavefunction, and $S(E)$ --- the astrophysical $S$-factor --- encodes the nuclear interaction part of the reaction.  The $S$-factor is an intrinsic property of the nuclear force; it is independent of the presence of a muon and of the specific treatment adopted for the collision dynamics.  In this work, we use the high-precision fitted $S(E)$ from Ref.~\cite{wang2026}.

\begin{figure}
    \centering
    \includegraphics[width=\linewidth]{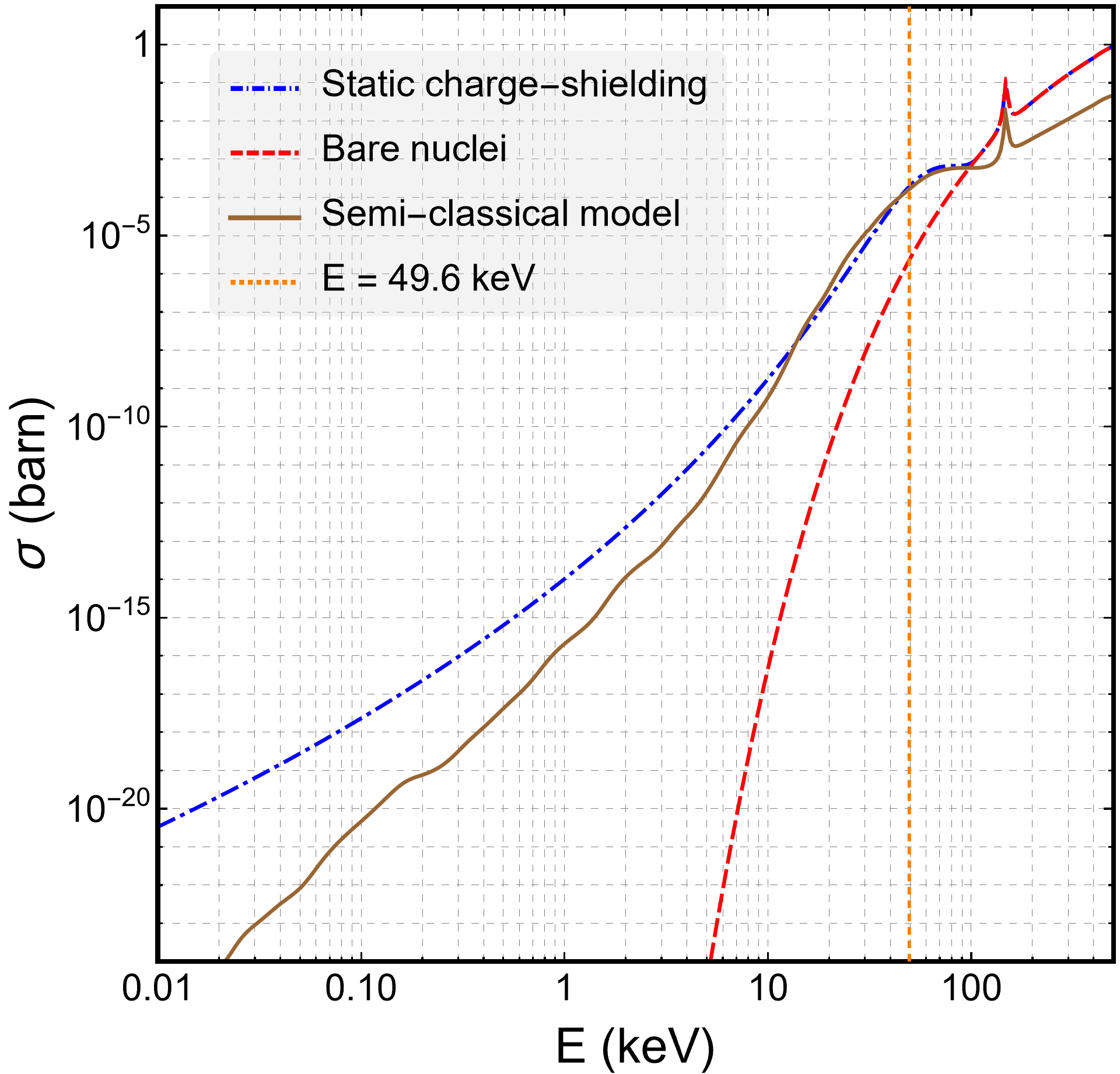}
    \caption{(Color online) Reaction cross-section $\sigma(E)$ as a function of incident kinetic energy $E$.}
    \label{fig:sigma}
\end{figure}

Although the $S$-factor is the same for all three cases considered here, the penetrability $P(E)$ differs markedly among the static charge-shielding treatment, our SC treatment, and the bare-nucleus picture.  Substituting the respective $P(E)$ into the Gamow formula yields the cross-sections displayed in Fig.~\ref{fig:sigma}. In the energy range $E \approx 0.01$--$10$~keV, the SC cross-section is noticeably lower than that predicted by the static charge-shielding treatment, yet it still exceeds the bare-nucleus cross-section by many orders of magnitude, demonstrating that muon enhancement is highly effective at low energies. For $E \gtrsim 100$~keV, the SC cross-section becomes slightly smaller than the bare-nucleus result, which is consistent with the analysis presented in the previous section.  Importantly, for the p--$^{11}$B reaction the muon does not enhance the first major resonance peak near $148$~keV; rather, our SC treatment indicates that the muon slightly suppresses the cross-section in this resonance region.

Finally, we compare the reactivities obtained from the three treatments. The reactivity is closely related to the cross-section and is given by~\cite{Rolfs1988}
\begin{equation}
\langle\sigma v\rangle
= \frac{(8/\pi)^{1/2}}{\mu_{pB}^{1/2}\,(k_{B}T)^{3/2}}
   \int_{0}^{\infty} \sigma E\,
   \exp\!\bigl(-E/k_{B}T\bigr)\,\mathrm{d}E,
\end{equation}
where $\mu_{pB}$ is the reduced mass of the p--$^{11}$B system, $k_{B}$ is the Boltzmann constant, $T$ is the temperature, and $\sigma$ is the reaction cross-section.

The calculated reactivities for selected temperatures are listed in Table~\ref{tab:reactivity}. As in the case of the cross-section, at temperatures below $10$~keV the SC treatment yields a reactivity somewhat lower than that of the static charge-shielding treatment, but both values exceed the bare-nucleus result by several orders of magnitude. Moreover, the lower the temperature, the stronger the catalytic enhancement provided by the muon. At higher temperatures, however, the influence of the muon becomes increasingly marginal. In the range $100$--$500$~keV, the static screening treatment gives reactivities nearly indistinguishable from the bare-nucleus case, while our SC treatment predicts reactivities slightly below the bare-nucleus values. For magnetic confinement p-$^{11}$B fusion, the energy window of interest for potential net energy gain lies approximately between $200$ and $300$~keV~\cite{wang2026}. Therefore, as pointed out in Ref.~\cite{Wang:2026zuj}, although the muon does not appreciably enhance the fusion power density under magnetic-confinement conditions, it dramatically alters the reaction landscape in the low-energy regime. When combined with techniques such as intense laser fields or high-density compression, this effect may offer a possible route toward low-energy auxiliary ignition of p-$^{11}$B fusion.

\begin{table}
\centering
\caption{Reactivity $\langle\sigma v\rangle$ ($10^{-22}m^3/s$) for the three treatments}
\label{tab:rates}
\label{tab:reactivity}
\begin{tabular}{c c c c}
\hline
$T$ (keV) \ & \ Bare nuclei \ & \ Static treatment \ & \ SC treatment\ \\
\hline
0.10  & $1.120\times10^{-42}$ & $2.893\times10^{-17}$ & $3.016\times10^{-19}$ \\
0.18  & $5.961\times10^{-34}$ & $4.886\times10^{-16}$ & $1.191\times10^{-17}$ \\
0.34  & $7.418\times10^{-27}$ & $1.102\times10^{-14}$ & $4.933\times10^{-16}$ \\
0.62  & $4.299\times10^{-21}$ & $3.630\times10^{-13}$ & $4.158\times10^{-14}$ \\
1.14  & $2.038\times10^{-16}$ & $1.968\times10^{-11}$ & $1.136\times10^{-11}$ \\
2.09  & $1.255\times10^{-12}$ & $1.913\times10^{-9}$  & $3.281\times10^{-9}$  \\
3.85  & $1.465\times10^{-9}$  & $2.272\times10^{-7}$  & $3.422\times10^{-7}$  \\
7.07  & $4.578\times10^{-7}$  & $1.120\times10^{-5}$  & $1.132\times10^{-5}$  \\
12.99 & $7.276\times10^{-5}$  & $1.931\times10^{-4}$  & $1.311\times10^{-4}$  \\
23.87 & $3.150\times10^{-3}$  & $3.536\times10^{-3}$  & $8.866\times10^{-4}$  \\
43.87 & $4.540\times10^{-2}$  & $4.591\times10^{-2}$  & $6.263\times10^{-3}$  \\
80.60 & $4.126\times10^{-1}$  & $4.118\times10^{-1}$  & $1.230\times10^{-1}$  \\
148.10 & $1.740\times10^{0}$  & $1.709\times10^{0}$  & $9.266\times10^{-1}$  \\
272.12 & $3.450\times10^{0}$  & $3.216\times10^{0}$  & $2.256\times10^{0}$  \\
500.00 & $4.698\times10^{0}$  & $3.685\times10^{0}$  & $2.964\times10^{0}$  \\
\hline
\end{tabular}
\end{table}

\section{Conclusion}
\label{conclusion}

In this work, we introduced a three-body semi-classical treatment to investigate muon-enhanced proton-boron-11 (p-$^{11}$B) fusion in a kinetic scenario, where a muonic hydrogen atom p$\mu$ is first formed and subsequently bombarded by a $^{11}$B nucleus.  This approach goes beyond the previously proposed static charge-shielding treatment by considering the p$\mu$-$^{11}$B system as a genuine three-body problem.  The classical turning point between the proton and the $^{11}$B is determined statistically via the classical trajectory Monte Carlo (CTMC) method, with the initial phase-space distribution of the p$\mu$ atom sampled from the microcanonical ensemble at the ground-state energy.  The tunneling probability is evaluated using the WKB approximation supplemented by the Airy-function connection technique, and the reaction cross-section is obtained from the Gamow form.

Our CTMC results confirm that the presence of a muon can enhance the low-energy p-$^{11}$B reaction cross-section by several orders of magnitude compared with the bare-nucleus case.  A detailed comparison with the static charge-shielding treatment reveals that the two approaches agree well in an intermediate energy window ($E \approx 1$--$50$~keV) but deviate systematically at both lower and higher energies.  These deviations are traced to effects that are naturally captured by the three-body CTMC simulation but are absent from the static screening picture: at low energies, the internal motion of the p$\mu$ atom introduces a small impact parameter that increases the turning-point distance; at high energies, the incident $^{11}$B velocity becomes comparable to the muon orbital velocity, so that the muon no longer acts as a stationary screening cloud and can even attract the $^{11}$B nucleus, thereby increasing $r_{\min}$.

The resulting reaction cross-section is largest at the lowest incident energies and decreases as the energy increases.  While the muon does not enhance the first major resonance peak near $148$~keV --- and in fact slightly suppresses it --- the cross-section in the sub-$100$~keV regime remains dramatically higher than that of bare nuclei.  The reactivity calculated from the cross-section shows qualitatively similar behavior: a strong catalytic enhancement at temperatures below $\sim 10$~keV, which gradually diminishes at higher temperatures and becomes negligible in the energy window relevant for magnetic-confinement fusion ($200$--$300$~keV).

Taken together, these results demonstrate that muon enhancement  profoundly reshapes the low-energy reaction landscape of p-$^{11}$B fusion.  Although it does not appreciably increase the fusion power density under typical magnetic-confinement conditions, the substantial enhancement of the cross-section and reactivity at sub-$100$~keV energies suggests that the p$\mu$-$^{11}$B channel may serve as a viable route for lowering the ignition threshold.  When combined with techniques such as intense laser fields or high-density compression, this mechanism may offer a promising pathway toward low-energy auxiliary ignition of aneutronic p-$^{11}$B fusion.

\begin{acknowledgments}
This work was supported by Natural Science Foundation of Jiangsu Province (grant no. BK20220122); China Postdoctoral Science Foundation (grant no. 2024M751369); National Natural Science Foundation of China (grant no. 12233002), and Jiangsu Funding Program for Excellent Postdoctoral Talent. Use of the computing resources at \href{www.syli.cloud}{Jiangsu Xilixi Technology} is gratefully acknowledged.
\end{acknowledgments}


\nocite{*}
\bibliographystyle{apsrev4-1}
\bibliography{refs} 

@article{Last2011,
    author = {Last, I. and Ron, S. and Jortner, J.},
    author1 = {Last, Isidore and Ron, Shlomo and Jortner, Joshua},
    title = {Aneutronic H + ${}^{11}$B Nuclear Fusion Driven by Coulomb Explosion of Hydrogen Nanodroplets},
    journal = {Phys. Rev. A},
    volume = {83},
    number = {4},
    pages = {043202},
    year = {2011},
    doi = {10.1103/PhysRevA.83.043202}
}

@article{Sciscio2025,
    author = {Scisci{\`o}, M. and Petringa, G. and Zhu, Z. and Rodrigues, M. R. D. and Alonzo, M. and Andreoli, P. L. and Filippi, F. and Consoli, F. and Huault, M. and Raffestin, D. and Molloy, D. and Larreur, H. and Singappuli, D. and Carriere, T. and Verona, C. and Nicolai, P. and McNamee, A. and Ehret, M. and Filippov, E. and others},
    title = {Laser-Initiated p--${}^{11}$B Fusion Reactions in Petawatt High-Repetition-Rate Laser Facilities},
    journal = {Matter Radiat. Extremes},
    volume = {10},
    number = {3},
    pages = {037401},
    year = {2025},
    doi = {10.1063/5.0241993}
}

@article{Liu2024,
    author = {Liu, M. and Xie, H. and Wang, Y. and Dong, J. and Feng, K. and Gu, X. and Huang, X. and Jiang, X. and Li, Y. and Li, Z. and Liu, B. and Liu, W. and Luo, D. and Peng, Y.-K. M. and Shi, Y. and Song, S. and Song, X. and Sun, T. and Tan, M. and Zhao, H.},
    title = {{ENN}'s Roadmap for Proton-Boron Fusion Based on Spherical Torus},
    journal = {Phys. Plasmas},
    volume = {31},
    number = {6},
    pages = {062701},
    year = {2024},
    doi = {10.1063/5.0199112}
}

@article{Beckman1953,
    author = {Beckman, O. and Huus, T. and Zupancic, C.},
    title = {Proton Bombardment of ${}^{11}$B},
    journal = {Phys. Rev.},
    volume = {91},
    pages = {606},
    year = {1953},
    doi = {10.1103/PhysRev.91.606}
}

@article{Moreau1977,
    author = {Moreau, D. C.},
    title = {Potentiality of the Proton-Boron Fuel for Controlled Thermonuclear Fusion},
    journal = {Nucl. Fusion},
    volume = {17},
    number = {1},
    pages = {13--20},
    year = {1977},
    doi = {10.1088/0029-5515/17/1/002}
}

@article{Mazzucconi2025,
    author = {Mazzucconi, D. and Bellotti, E. S. and Vavassori, D. and Dellasega, D. and Agosteo, S. and Passoni, M. and Pola, A. and Bortot, D.},
    title = {Evaluation of the p--${}^{11}$B Reaction Cross Section through a Silicon Telescope in the 0.3--4.7 MeV Range},
    journal = {Eur. Phys. J. A},
    volume = {61},
    number = {5},
    pages = {114},
    year = {2025},
    doi = {10.1140/epja/s10050-025-01589-3}
}

@incollection{Dawson1981,
    author = {Dawson, J. M.},
    title = {Advanced Fusion Reactors},
    booktitle = {Fusion},
    editor = {Teller, E.},
    volume = {1},
    pages1 = {453},
    publisher = {Academic Press},
    address = {New York},
    year = {1981},
    note = {Part B},
    c={Dawson, J. Fusion, (ed. Teller, E.) Part B, Vol. 1 (Academic Press, 1981).}
}

@article{Ogawa_2024,
    title = {Demonstration of aneutronic p-11B reaction in a magnetic confinement device},
    volume = {64},
    url = {https://iopscience.iop.org/article/10.1088/1741-4326/ad6615},
    doi = {10.1088/1741-4326/ad6615},
    journal = {Nucl. Fusion},
    publisher = {IOP Publishing},
    author = {Ogawa, K. and others},
    year = {2024},
    pages = {096028}
}

@article{Tentori2023,
    author = {Tentori, A. and Belloni, F.},
    title = {Revisiting p--${}^{11}$B Fusion Cross Section and Reactivity, and Their Analytic Approximations},
    journal = {Nucl. Fusion},
    volume = {63},
    number = {8},
    pages = {086001},
    year = {2023},
    doi = {10.1088/1741-4326/acda4b}
}

@article{Putvinski2019,
    author = {Putvinski, S. V. and Ryutov, D. D. and Yushmanov, P. N.},
    title = {Fusion Reactivity of the p${}^{11}$B Plasma Revisited},
    journal = {Nucl. Fusion},
    volume = {59},
    number = {7},
    pages = {076018},
    year = {2019},
    doi = {10.1088/1741-4326/ab1a60}
}

@article{Hartouni_2022,
    title = {Evidence for suprathermal ion distribution in burning plasmas},
    volume = {19},
    issn = {1745-2481},
    url = {https://doi.org/10.1038/s41567-022-01809-3},
    doi = {10.1038/s41567-022-01809-3},
    number = {1},
    journal = {Nat. Phys.},
    publisher = {Springer Science and Business Media LLC},
    author = {Hartouni, E. P. and Moore, A. S. and Crilly, A. J. and Appelbe, B. D. and Amendt, P. A. and Baker, K. L. and Casey, D. T. and Clark, D. S. and Döppner, T. and Eckart, M. J. and Field, J. E. and Gatu-Johnson, M. and Grim, G. P. and Hatarik, R. and Jeet, J. and Kerr, S. M. and Kilkenny, J. and Kritcher, A. L. and Meaney, K. D. and {others}},
    year = {2022},
    month = {Jul},
    pages = {72--77}
}

@article{Wang:2026zuj,
    author = "Wang, Hong-Yi and Li, Yu-Qi and Wu, Qian and Cui, Zhu-Fang",
    title = "{A novel approach to proton-boron-11 fusion}",
    eprint = "2604.18928",
    archivePrefix = "arXiv",
    primaryClass = "nucl-th",
    month = "4",
    year = "2026"
}

@article{wang2026,
    author = "Wang, Hong-Yi and Li, Yu-Qi and Wu, Qian and Cui, Zhu-Fang",
    title = "{Revisiting p-$^{11}$B Fusion: Updated Cross-sections, Reactivity, and Energy Balance}",
    eprint = "2601.00241",
    archivePrefix = "arXiv",
    primaryClass = "nucl-th",
    month = "1",
    year = "2026"
}

@article{Wu:2025bnm,
    author = "Wu, Qian and Cui, Zhu-Fang and Kamimura, Masayasu",
    title = "{Reaction processes of muon-catalyzed fusion in the muonic molecule dd{\ensuremath{\mu}} studied with the tractable T-matrix model}",
    eprint = "2508.12783",
    archivePrefix = "arXiv",
    primaryClass = "nucl-th",
    doi = "10.1103/3ddg-3p73",
    journal = "Phys. Rev. C",
    volume = "113",
    number = "3",
    pages = "034603",
    year = "2026"
}

@article{Wu2024,
    author = {Wu, Qian and Kamimura, Masayasu},
    title = {Tractable T-Matrix Model for Reaction Processes in Muon-Catalyzed Fusion (dt$\mu$)J=v=0$\rightarrow$$\alpha$+n+$\mu$+17.6 MeV or ($\alpha$$\mu$)nl+n+17.6 MeV},
    eprint1 = {2401.17358},
    archiveprefix1 = {arXiv},
    primaryclass1 = {nucl-th},
    doi = {10.1103/PhysRevC.109.054625},
    journal = {Phys. Rev. C},
    volume = {109},
    number = {5},
    pages = {054625},
    year = {2024}
}

@article{Froelich1992,
author = {P. Froelich},
title = {Muon catalysed fusion Chemical confinement of nuclei within the muonic molecule dt$\mu$},
journal1 = {Advances in Physics},
journal = {Adv. Phys.},
volume = {41},
number = {5},
pages = {405--508},
year = {1992},
publisher = {Taylor \& Francis},
doi1 = {10.1080/00018739200101533},
URL = {https://doi.org/10.1080/00018739200101533},
eprint1 = {https://doi.org/10.1080/00018739200101533}
}

@ARTICLE{Rafelski1987,
       author = {{Rafelski}, Johann and {Jones}, Steven E.},
        title = "{Cold nuclear fusion}",
      journal = {Sci. Am.},
         year = 1987,
        month = jul,
       volume = {257},
        pages = {84-89},
          doi = {10.1038/scientificamerican0787-84},
       adsurl = {https://ui.adsabs.harvard.edu/abs/1987SciAm.257a..84R}
}

@article{Koshigiri1984,
    author = {Koshigiri, Kunio and Ohtsubo, Hisao and Morita, Masato},
    title = {Muon Capture in 11B Muonic Atom and Hyperfine Spin Dependence: },
    journal = {Prog. Theor. Phys.},
    volume = {71},
    number = {6},
    pages = {1293-1302},
    year = {1984},
    month = {06},
    issn = {0033-068X},
    doi = {10.1143/PTP.71.1293},
    url = {https://doi.org/10.1143/PTP.71.1293},
    eprint1 = {https://academic.oup.com/ptp/article-pdf/71/6/1293/5312546/71-6-1293.pdf}
}

@article{PhysRevC.65.025503,
  title = {Muon capture by ${}^{11}\mathrm{B}$ and the hyperfine effect},
  author = {Wiaux, V. and Prieels, R. and Deutsch, J. and Govaerts, J. and Brudanin, V. and Egorov, V. and Petitjean, C. and Tru\"ol, P.},
  journal = {Phys. Rev. C},
  volume = {65},
  issue = {2},
  pages = {025503},
  numpages = {8},
  year = {2002},
  month = {Jan},
  publisher = {American Physical Society},
  doi = {10.1103/PhysRevC.65.025503},
  url = {https://link.aps.org/doi/10.1103/PhysRevC.65.025503}
}

@inproceedings{Schaller1993,
    editor = {Lukas A. Schaller and Claude Petitjean},
    year = {1993},
    booktitle = {Muonic Atoms and Molecules},
    series = {Monte Verita},
    doi={https://doi.org/10.1007/978-3-0348-7271-3},
    publisher = {Birkhäuser Basel},
    edition = {1}
}

@book{liu2013classical,
  author    = {Liu, J.},
  title     = {{Classical Trajectory Perspective of Atomic Ionization in Strong Laser Fields: Semiclassical Modeling}},
  publisher = {Springer},
  address   = {New York},
  year      = {2013}
}

@incollection{percival1975theory,
  author    = {Percival, I. C. and Richards, D.},
  title     = {The Theory of Collisions between Charged Particles and Highly Excited Atoms},
  booktitle = {Advances in Atomic and Molecular Physics},
  editor    = {Bates, D. R. and Bederson, Benjamin},
  volume    = {11},
  pages     = {1--82},
  publisher = {Academic Press},
  address   = {New York},
  year      = {1975},
  doi       = {10.1016/S0065-2199(08)60028-7}
}

@article{abrines1966classical,
  author  = {Abrines, R. and Percival, I. C.},
  title   = {Classical Theory of Charge Transfer and Ionization of Hydrogen Atoms by Protons},
  journal = {Proc. Phys. Soc.},
  volume  = {88},
  number  = {4},
  pages   = {861--872},
  year    = {1966},
  doi     = {10.1088/0370-1328/88/4/309}
}

@article{abrines1966classical2,
  author  = {Abrines, R. and Percival, I. C. and Valentine, N. A.},
  title   = {Classical Cross Sections for Ionization of Hydrogen Atoms by Electrons},
  journal = {Proc. Phys. Soc.},
  volume  = {89},
  number  = {3},
  pages   = {515--523},
  year    = {1966},
  doi     = {10.1088/0370-1328/89/3/307}
}

@article{verlet1967computer,
  author    = {Verlet, Loup},
  title     = {Computer ``Experiments'' on Classical Fluids. I. Thermodynamical Properties of Lennard-Jones Molecules},
  journal   = {Phys. Rev.},
  volume    = {159},
  number    = {1},
  pages     = {98--103},
  year      = {1967},
  doi       = {10.1103/PhysRev.159.98}
}

@article{swope1982computer,
  author    = {Swope, William C. and Andersen, Hans C. and Berens, Peter H. and Wilson, Kent R.},
  title     = {A Computer Simulation Method for the Calculation of Equilibrium Constants for the Formation of Physical Clusters of Molecules: Application to Small Water Clusters},
  journal   = {J. Chem. Phys.},
  volume    = {76},
  number    = {1},
  pages     = {637--649},
  year      = {1982},
  doi       = {10.1063/1.442716}
}

@online{IAEANuclearData,
    author    = {{IAEA Nuclear Data Services}},
    title     = {Table of Nuclear Charge Radii},
    year      = {2023},
    url       = {https://www-nds.iaea.org/radii/},
    note1      = {Accessed: 2023-10-27},
    urldate   = {2023-10-27}
}

@article{cohen1982comment,
  author  = {Cohen, James S.},
  title   = {Comment on the Classical-Trajectory Monte Carlo Method for Ion-Atom Collisions},
  journal = {Phys. Rev. A},
  volume  = {26},
  number  = {5},
  pages   = {3008--3010},
  year    = {1982},
  doi     = {10.1103/PhysRevA.26.3008}
}

@article{leopold1979ionisation,
  author  = {Leopold, J. G. and Percival, I. C.},
  title   = {Ionisation of Highly Excited Atoms by Electric Fields. {III}. Microwave Ionisation and Excitation},
  journal = {J. Phys. B: Atom. Mol. Phys.},
  volume  = {12},
  number  = {5},
  pages   = {709--721},
  year    = {1979},
  doi     = {10.1088/0022-3700/12/5/016}
}

@book{LandauLifshitz1977,
    author    = {L. D. Landau and E. M. Lifshitz},
    title     = {Quantum Mechanics: Non-Relativistic Theory},
    edition   = {3rd},
    series    = {Course of Theoretical Physics, Vol. 3},
    publisher = {Pergamon Press},
    year      = {1977}
   
}

@book{Olver1974,
    author    = {F. W. J. Olver},
    title     = {Asymptotics and Special Functions},
    publisher = {Academic Press},
    year      = {1974}
}

@book{AbramowitzStegun1964,
    author    = {Milton Abramowitz and Irene A. Stegun},
    title     = {Handbook of Mathematical Functions with Formulas, Graphs, and Mathematical Tables},
    publisher = {National Bureau of Standards},
    year      = {1964}
}

@book{Rolfs1988,
    author = {Rolfs, Claus E. and Rodney, William S.},
    title = {Cauldrons in the Cosmos: Nuclear Astrophysics},
    publisher = {University of Chicago Press},
    year = {1988},
    address = {Chicago},
    isbn = {0-226-72456-5}
}
\end{document}